\documentclass[%
 superscriptaddress,
 preprint,
 aps,
 prb,
]{revtex4-1}

\usepackage{graphicx}% Include figure files
\begin{document}

%\preprint{}

\title{Ferromagnetic Instability in a Doped Band-Gap Semiconductor FeGa$_{3}$}

\author{K. Umeo}
\email[]{kumeo@sci.hiroshima-u.ac.jp}
\affiliation{Cryogenics and Instrumental Analysis Division, N-BARD, Hiroshima University, Higashi-Hiroshima 739-8526, Japan}
\affiliation{Department of Quantum matter, AdSM, Hiroshima University, Higashi-Hiroshima 739-8530, Japan}
\author{Y. Hadano}
\affiliation{Department of Quantum matter, AdSM, Hiroshima University, Higashi-Hiroshima 739-8530, Japan}
\author{S. Narazu}
\affiliation{Department of Quantum matter, AdSM, Hiroshima University, Higashi-Hiroshima 739-8530, Japan}
\author{T. Onimaru}
\affiliation{Department of Quantum matter, AdSM, Hiroshima University, Higashi-Hiroshima 739-8530, Japan}
\author{M. A. Avila}
\affiliation{Centro de Ci\^{e}ncias Naturais e Humanas, Universidade Federal do ABC, Santo Andr\'{e} - SP, 09210-971, Brazil}
\author{T. Takabatake}
\affiliation{Department of Quantum matter, AdSM, Hiroshima University, Higashi-Hiroshima 739-8530, Japan}
\affiliation{Institute for Advanced Materials Research, Hiroshima University, Higashi-Hiroshima 739-8530, Japan}

\date{\today}

\begin{abstract}

We report the effects of electron doping on the ground state of a 
diamagnetic semiconductor FeGa$_{3}$ with a band gap of $0.5$ eV. By means of 
electrical resistivity, magnetization and specific heat measurements we have 
found that gradual substitution of Ge for Ga in FeGa$_{3-y}$Ge$_{y}$ 
yields metallic conduction at a very small level of $y = 0.006$, then induces 
weak ferromagnetic (FM) order at $y = 0.13$ with a spontaneous moment of 0.1 
$\mu_{B}$/Fe and a Curie temperature $T_{C}= 3.3$ K, which continues 
increasing to $T_{C} = 75$ K as doping reaches $y = 0.41$. The emergence of 
the FM state is accompanied by quantum critical behavior as observed in the 
specific heat, $C/T \propto -$ln$T$, and in the magnetic susceptibility, $M/B \propto 
 T^{-4/3}$. At $y= 0.09$, the specific heat divided by 
temperature $C/T$ reaches a large value of 70 mJ/K$^{2}$molFe, twice as large as 
that reported on FeSi$_{1-x}$Ge$_{x}$ for $x_{c}= 0.37$ and 
Fe$_{1-x}$Co$_{x}$Sb$_{2}$ for $x_{c}=0.3$ at their respective FM 
quantum critical points. The critical concentration $y_{c}=0.13$ in 
FeGa$_{3-y}$Ge$_{y}$ is quite small, despite the fact that its band gap 
is one order of magnitude larger than those in FeSi and FeSb$_{2}$. In 
contrast, no FM state emerges by substituting Co for Fe in 
Fe$_{1-x}$Co$_{x}$Ga$_{3}$ in the whole range $0 \leq x \leq  1$, 
although both types of substitution should dope electrons into FeGa$_{3}$. 
The FM instability found in FeGa$_{3-y}$Ge$_{y}$ indicates that strong 
electron correlations are induced by the disturbance of the Fe $3d$ - Ga $4p$ 
hybridization.

\end{abstract}

\maketitle

%\tableofcontents

\section{\label{sec:level1}Introduction}
Iron- and ruthenium-based semiconductors with band gaps of the order of 0.1 
eV such as FeSi,\cite{VJ01, JB02, ZS03, DM04, VI05, SP06, BB07, VI08, TS09, SY10, YN11, MA12, HY13, DP14, JI15} FeSb$_{2}$,\cite{CP16, CP17, AP18, AB19, RH20, RH21, AB22} FeGa$_{3}$,\cite{CD23, UH24, YA25, YI26, NT27, YH28, EM29, ZP30, NH31, MA32, VG33} 
Fe$_{2}$VAl,\cite{YN34} RuAl$_{2}$,\cite{DN35} RuGa$_{3}$,\cite{UH24} and 
RuIn$_{3}$,\cite{DB36} have attracted considerable attention because of their 
unusual transport and magnetic behaviors. These compounds have been 
intensively studied not only as candidate thermoelectric materials, but also 
from an academic interest in the mechanism of the gap formation, which has 
been discussed in the context of strong correlations involving $3d$ or $4d$ bands, 
analogous to $4f$ bands in rare-earth-based Kondo semiconductors. In typical 
$4f$ Kondo semiconductors such as YbB$_{12}$ and Ce$_{3}$Pt$_{3}$Bi$_{4}$, a 
small gap of about $0.02$ eV is formed by the hybridization of localized $4f$ 
states with the conduction bands.\cite{TT37} Kondo semiconductors are 
distinguished from band-gap semiconductors on the following points: (i) The 
gap gradually disappears upon heating to a temperature which is lower than 
the gap energy, as observed in the temperature dependence of optical 
conductivity for FeSi and FeSb$_{2}$.\cite{ZS03, SP06, AP18} (ii) The gap is strongly 
suppressed by substituting both the magnetic ion site and the ligand site at 
a low level. Thereby, the magnetization and the electronic specific heat 
coefficient are largely enhanced. This enhancement is observed in 
Fe$_{1-x}$Co$_{x}$Si,\cite{JB02, YN11} FeSi$_{1-x}$Ge$_{x}$,\cite{SY10} 
Fe$_{1-x}$Co$_{x}$Sb$_{2}$,\cite{RH20, RH21} and 
FeSb$_{2-x}$Sn$_{x}$.\cite{AB19} Recently, however, the above physical 
properties of FeSi and FeSb$_{2}$ have been explained by a minimum model of 
covalent insulator within a single-site dynamical mean-field 
approximation.\cite{JK38} Furthermore, the electronic structure of FeSi measured 
by photoemission experiments has no distinct features relevant to a Kondo 
picture, but is qualitatively explained within the band calculations by the 
density functional theory without many-body effects,\cite{HY13, DZ39, MK40} 
Therefore, it remains an issue whether FeSi and FeSb$_{2}$ are Kondo or 
usual semiconductors.

 FeSi and FeSb$_{2}$ are nearly ferromagnetic semiconductors. In spite of 
the absence of magnetic order in both FeSi and CoSi, their mixed system 
Fe$_{1-x}$Co$_{x}$Si exhibits magnetic order in the range 0.05 
\textless $x$ \textless 0.8.\cite{JB02, YN11} Small angle neutron scattering 
experiments have revealed a helical spin magnetic structure with a long 
period of more than 300 {\AA}.\cite{JB02} This magnetic structure is realized by 
the Dzyaloshinsky-Moriya interaction as found in B20 crystal structures 
without inversion center. By applying magnetic fields, the helical structure 
easily transforms to the FM one. Moreover, FeSi$_{1-x}$Ge$_{x}$ ($x \ge  
0.37$) and Fe$_{1-x}$Co$_{x}$Sb$_{2}$ ($0.2 \le x \textless 0.5$) also 
present the emergence of ferromagnetism.\cite{SY10, RH20, RH21} According to the 
local density approximation plus on-site Coulomb repulsion correction 
method, the semiconducting states in FeSi and FeSb$_{2}$ are close in energy 
to a FM and metallic state.\cite{VI05, AV41} Thereby, local Coulomb repulsions 
$U$ of 3.7 eV and 2.6 eV were obtained for FeSi and FeSb$_{2}$, respectively.

FeGa$_{3}$ crystallizes into a tetragonal structure with space group 
$P$4$_{2}$/\textit{mnm}. A narrow $d $(Fe) - $p $(Ga) hybridization band-gap $E_{g} = 0.3-0.5 $eV is 
expected from the band structure calculations based on the 
density-functional theory within the local density approximation.\cite{UH24, YI26} It is consistent with the observed gap of $0.25-0.47$ eV (Refs. 25, 27, 
28, 32) for FeGa$_{3}$. This value is one order of magnitude larger than 
that in FeSi (Ref. 4) and FeSb$_{2}$,\cite{CP16} whose gaps are 0.08 and 0.02 
eV, respectively. In FeGa$_{3}$, the absence of a significant 
impurity-induced density of states at the Fermi level $E_{F}$ is indicated by 
an extremely small $\gamma$ value of 0.03 mJ/K$^{2}$mol.\cite{YH28} These facts 
suggest that correlation effects or the nature of the Kondo semiconductor in 
FeGa$_{3}$ are weaker than in FeSi and FeSb$_{2}$. This weak correlation 
effect in FeGa$_{3}$ manifests itself by the absence of a sharp peak at the 
valence band maximum just below $E_{F}$, as found in recent photoemission 
spectra.\cite{MA32}  The magnetic susceptibility shows diamagnetism below room 
temperature, and it increases exponentially with temperature above 500 
K.\cite{NT27, YH28} Recently, it has been reported that Co substitution for Fe 
in Fe$_{1-x}$Co$_{x}$Ga$_{3}$ ($x = 0.05$) induces a crossover from the 
semiconducting state to a metallic state with weakly coupled local 
moments.\cite{EM29} 

In order to investigate the mechanism of metallization and emergence of 
ferromagnetism induced by electron doping in FeGa$_{3}$, we have synthesized 
$3d$ electron doped Fe$_{1-x}$Co$_{x}$Ga$_{3}$ samples and $4p$ electron 
doped FeGa$_{3-y}$Ge$_{y}$ samples and measured the electrical 
resistivity $\rho$, specific heat $C$, and magnetization $M$. Our results 
demonstrate a doping-induced semiconductor-metal transition in both systems, 
but weak FM state only in FeGa$_{3-y}$Ge$_{y}$ for $y \ge  0.13$. We 
will discuss how the doping effects in the FeGa$_{3}$ system differ from 
those in the FeSi and FeSb$_{2}$ systems.

\section{Experimental details}
Single crystals of Fe$_{1-x}$Co$_{x}$Ga$_{3}$ and 
FeGa$_{3-y}$Ge$_{y}$ were grown by a Ga self-flux method. Mixtures of 
high purity elements in compositions Fe : Co : Ga $= 1-X $: $X $: 9 ($0 \leq X 
\le  1$) and Fe : Ga : Ge $=$ 1 : 8.5 : $Y$ ($0.01 \le Y  \le  3$) were 
sealed in evacuated silica ampoules. The ampoules were heated to 1100 
$^{\circ}$C and cooled over 150 hours to 500 $^{\circ}$C , at which point 
the molten Ga flux was separated by decanting. The crystal compositions were 
determined by electron-probe microanalysis (EPMA) using a JEOL JXA-8200 
analyzer. The effective Co doping levels in the crystals were found to 
roughly agree with the nominal composition $X$, whereas a maximum effective Ge 
doping of $y = 0.41$ results for an initial composition $Y = 3$. X-ray 
diffraction patterns of powdered samples confirmed that all alloys for 0 
$\le x \le $ 1 and $y \le  0.41$ crystallized in the FeGa$_{3}$-type 
structure. No impurity phases in the single crystals were found by x-ray 
diffraction nor EPMA. The lattice parameters $a$ and $c$, and the unit cell volume 
$V$ are plotted in Fig. 1. The values of $a = 6.262$ (6.240) and $c = 6.556$ 
(6.439) {\AA} of FeGa$_{3}$ (CoGa$_{3})$ are in good agreement with reported 
values.$^{24,\, 43}$ For Fe$_{1-x}$Co$_{x}$Ga$_{3}$, both $a$ and $c$ 
parameters decrease monotonically with increasing $x$ from 0 to 1, following 
Vegard's law. The $V(x = 1$) is 2.5 {\%} smaller than $V(x = 0$). For 
FeGa$_{3-y}$Ge$_{y}$, the $a$ value increases with increasing $y$, whereas 
the $c$ value decreases. As a result, $V(y = 0.41$) is only 1{\%} smaller than 
$V(y = 0$). 

Resistivity measurements were performed on a home-built system using a 
standard four-probe AC method, in the temperature range of $3-380$ K provided 
by a Gifford-McMahon type refrigerator. The magnetization $M$ was measured 
under ambient pressure as well as applied pressures up to 2.21 GPa by using 
a SQUID magnetometer (Quantum Design MPMS) from 2 to 350 K and in magnetic 
fields up to 5 T. To measure $M$ down to 0.35 K, we adopted a capacitive 
Faraday method using a high resolution capacitive force-sensing device 
installed in a $^{3}$He refrigerator.\cite{TS44} The specific heat $C$ from 0.3 to 
300 K was measured by a relaxation method on a Quantum Design PPMS.

\section{Results}
Figures 2(a) and (b) show the temperature dependence of $\rho$ for 
Fe$_{1-x}$Co$_{x}$Ga$_{3}$ and FeGa$_{3-y}$Ge$_{y}$, respectively. 
For Fe$_{1-x}$Co$_{x}$Ga$_{3}$, the data are normalized by the $\rho$ 
value at 380 K. The $\rho (T)$ data for $x = 0$ shown in the inset of Fig. 2(a) 
exhibits upturns in the temperature ranges of $T \textgreater 260$ K and $T 
\textless 50$ K, which are attributed to intrinsic response due to the band 
gap of 0.5 eV, and extrinsic one due to the impurity donors, 
respectively.\cite{YH28} The $\rho (T)/\rho_{380}$ for $x = 0.02$ increases with 
decreasing temperature in the entire temperature range. With increasing $x$, 
the upturn in $\rho (T)/\rho_{380\, }$is suppressed and $\rho 
(T)/\rho_{380}$ for $x \ge  0.23$ shows metallic behavior. On the other 
hand, the substitution of Ge for Ga in FeGa$_{3-y}$Ge$_{y}$ at a very 
small level of $y = 0.006$ yields metallic conduction. It should be recalled 
that for Fe$_{1-x}$Co$_{x}$Si and FeSi$_{1-x}$Ge$_{x}$, the 
semiconductor-metal transition occurs at high levels of substitution $x = 0.6$ 
and 0.25, respectively.\cite{SY10, YN11} Despite the fact that the band gap of 0.5 
eV for FeGa$_{3}$ is one order of magnitude larger than the one in FeSi, 
metallization occurs in FeGa$_{3-y}$Ge$_{y}$ at a much smaller doping 
level, suggesting that Ge substitution in FeGa$_{3}$ introduces drastic 
changes in the electronic state.

The temperature dependence of the magnetic susceptibility $M/B$ and its inverse 
$B/M$ for Fe$_{1-x}$Co$_{x}$Ga$_{3}$ are displayed in Figs. 3(a), (b), and 
(c). The diamagnetic behavior for $x = 0$ and 1 suggests that the Fermi level 
lies in the energy gap. The $M/B(T)$ for $0.1 \le  x \le  0.72$ shows Curie-Weiss 
paramagnetic behavior above 50 K. The negative value of the paramagnetic 
Curie temperature $\theta_{p}$ for $0.1 \le x \le  0.72$ implies that an 
antiferromagnetic interaction is dominant in this range. 

On the other hand, a ferromagnetic (FM) order occurs in 
FeGa$_{3-y}$Ge$_{y}$ for $y \ge  0.13$. As shown in Fig. 4, a 
spontaneous magnetic moment saturation $\mu_{s}$ is observed in the 
magnetization curves $M(B)$ for $y \ge  0.13$ at 2 K, and the value of $\mu 
_{s}$ increases with increasing $y$. However, the value of $\mu_{s}$ is 
significantly smaller than that of Fe metal, 2.22 $\mu_{B}$/Fe.\cite{SB45} 
Furthermore, the $M/B$ data as a function of temperature shows a ferromagnetic 
behavior for $y \ge  0.13$ as shown in Fig. 5. This FM transition should be a 
bulk property because $C(T)$ has a clear anomaly at the $T_{C}$ determined by the 
$M/B$ data, as shown in the inset of Fig.5. 

Figure 6 shows the temperature dependence of inverse magnetic susceptibility 
$B/M$ of FeGa$_{3-y}$Ge$_{y}$. For $y \ge  0.08$, the $B/M$ data follow the 
Curie-Weiss law. The value of $\theta_{p}$ for $y \le 0.09$ is negative, 
and changes to positive for $y \ge 0.13$. Both $\theta_{p}$ and Curie 
temperature $T_{C}$ as a function of $y$ are displayed in the upper panel of the 
inset of Fig. 6. The $T_{C}$ was estimated as the temperature where the 
extrapolation of $M(T)^{2}$ becomes zero. The increase in both $\theta_{p}$ 
and $T_{C}$ with increasing $y$ indicates that the FM interaction is enhanced by 
Ge doping.

In order to study the nature of ferromagnetism in FeGa$_{3-y}$Ge$_{y}$ 
for $y \ge 0.13$, the pressure dependence of $M$ has been measured. Figure 7 
shows the temperature dependence of $M/B$ for $y=0.34$ under various pressures $P$ 
and the inset shows the pressure dependence of $T_{C}$. It is found that 
$T_{C}$ decreases as $T_{C} \propto P^{3/4}$ which is predicted by the 
spin-fluctuation theory.\cite{AJ46} Furthermore, as shown in the lower panel of 
the inset of Fig. 6, the ratio of $\mu_{eff}$/$\mu_{s}$ is as high as 
4 -- 10. These findings suggest that FeGa$_{3-y}$Ge$_{y}$ 
for $y \ge 0.13$ is an itinerant weak ferromagnet. 

The specific heat divided by temperature, $C/T$, as a function of $T^{2}$ for 
Fe$_{1-x}$Co$_{x}$Ga$_{3}$ and FeGa$_{3-y}$Ge$_{y}$ is shown in 
Fig. 8. The $C/T$ data of FeGa$_{3-y}$Ge$_{y}$ for 0.05 $\le y \le 0.15$ 
displays an upturn below 5 K. The electronic specific-heat coefficient 
$\gamma $ was estimated by the extrapolation of the $C/T$ data to $T = 0$. The 
variations of $T_{C}(y)$ for FeGa$_{3-y}$Ge$_{y}$ and $\gamma $ ($x$ and 
$y)$ for Fe$_{1-x}$Co$_{x}$Ga$_{3}$ and FeGa$_{3-y}$Ge$_{y}$ are 
shown in Figs. 9 (a) and (b). It is worth noting that $\gamma (y)$ exhibits a 
sharp peak of 70 mJ/K$^{2}$mol at $y = 0.09$ near the critical concentration 
$y_{c}= 0.13$ where the ground state changes from a nonmagnetic state to 
a FM state, clearly contrasting with the almost flat behavior in $\gamma 
(x)$ for Fe$_{1-x}$Co$_{x}$Ga$_{3}$. The value of 70 mJ/K$^{2}$mol for 
$\gamma (y = 0.09$) is enhanced by a factor of 2300 compared to $\gamma (y =$ 
0) $= 0.03$ mJ/K$^{2}$mol, indicating the appearance of a heavy-fermion 
state in the vicinity of the FM instability.

The FM quantum critical behavior in $C/T$ and $M/B$ for FeGa$_{3-y}$Ge$_{y}$ ($y = 
0.09$) are evidenced in the plots in Fig. 10. The specific heat and magnetic 
susceptibility for $y = 0.09$ follow the functional forms of $C/T \propto  
-$ln$T$ and $M/B \propto T^{-4/3}$, which are predicted by the self-consistent 
renormalization (SCR) theory for FM spin fluctuations in three dimensional 
systems.\cite{TM47} These observations are consistent with the pressure 
dependence of $T_{C} \propto P^{3/4}$ in Fig. 7. On the other hand, as 
shown in Fig. 11, the $T$-linear dependence of $\rho (T)$ resistivity near the 
critical concentration of $y = 0.15$ is at variance with the $T^{5/3}$ 
dependence predicted by the SCR theory. The $\rho (T)$ data for $y = 0.08$ at 
$T \textless 30$ K obeys $T^{1.9}$, which indicates the recovery of the 
Fermi-liquid state. We will discuss the quantum critical behavior in 
FeGa$_{3-y}$Ge$_{y}$ in the next section.

\section{Discussions}
We now compare the doping effects on the electronic and magnetic states in 
Fe$_{1-x}$Co$_{x}$Ga$_{3}$ and FeGa$_{3-y}$Ge$_{y}$ with those in 
the FeSi and FeSb$_{2}$ systems. For Fe$_{1-x}$Co$_{x}$Ga$_{3}$, the 
semiconductor-metal transition occurs at $x = 0.23$, whereas no magnetically 
ordered state is induced in the whole range 0 $\le x \le  1$. The gradual 
and weak change of $\gamma (x)$ for Fe$_{1-x}$Co$_{x}$Ga$_{3}$ suggests 
that the band structure changes in the rigid-band frame. A similar situation 
has been observed in Fe$_{1-x}$Co$_{x}$Si, which exhibits a helical 
magnetically ordered state in the range 0.05 $\le x \textless 0.8$.\cite{YN11} A 
photoemission study on Fe$_{1-x}$Co$_{x}$Si revealed that the $x$ 
dependence of the band structure near the Fermi level is described by the 
rigid-band model.\cite{JY48} Therefore, the Stoner criterion can be applied to 
describe the magnetism of Fe$_{1-x}$Co$_{x}$Ga$_{3}$ and 
Fe$_{1-x}$Co$_{x}$Si. The criterion for the ferromagnetic state is 
given by the relation \textit{UD}($\varepsilon_{F}) \ge 1$, where $U$ and 
$D(\varepsilon_{F})$ are Coulomb repulsion and the density of states (DOS) 
at the Fermi level, respectively.\cite{SB45} From a photoemission spectroscopy 
study of FeGa$_{3}$, the magnitude of $U$ was estimated as 3 eV, which is 
comparable with 3.7 eV for FeSi.\cite{VI05, MA32} Therefore, the absence of a 
magnetically ordered state in Fe$_{1-x}$Co$_{x}$Ga$_{3}$ is a result of 
the fact that $D(\varepsilon_{F})$ at the bottom of the conduction band 
for Fe$_{1-x}$Co$_{x}$Ga$_{3}$ is smaller than that for 
Fe$_{1-x}$Co$_{x}$Si. 

On the other hand, for FeGa$_{3-y}$Ge$_{y}$, electron doping at a small 
level $y = 0.006$ already induces the semiconductor-metal transition. The Ga 
site substitution disturbs the $3d$-$4p$ hybridization, which should lead to a 
dramatic change in the electronic state. Higher doping for $y \ge  0.13$ 
yields a FM order. The doping induced FM state in the analogous system 
FeSi$_{1-x}$Ge$_{x}$ was explained by a mean-field slave-boson 
approach.\cite{SY10, VD49} Thereby, the key parameter driving the magnetic 
phases is ratio between the Coulomb repulsion $U$ and the hybridization of the 
localized-conduction electrons $V$. With increasing $U/V$, the paramagnetic ground 
state changes into an antiferromagnetic state and furthermore a FM 
state.\cite{VD49} For FeGa$_{3-y}$Ge$_{y}$, the disturbance of the ligand 
Ga/Ge site may lead to the suppression of the $d$-$p$ hybridization $V$, whereas $U$ in 
the Fe 3d shell would remain unchanged. Therefore, the Ga site substitution 
can yield the increase of $U/V$ and thus induce a FM ground state. On the other 
hand, for Fe$_{1-x}$Co$_{x}$Ga$_{3}$, the Fermi level shifts 
maintaining a rigid band, whereby $V$ does not change. Because $U/V$ is almost 
constant against $x$, no magnetic order is realized. Very recently, the 
experimental data for resistivity, specific heat and magnetization of 
FeSi$_{1-x}$Ge$_{x}$ have been explained by a minimal microscopic 
model.\cite{DP14, JI15} It is highly desirable to study whether this microscopic 
model is applicable for FeGa$_{3-y}$Ge$_{y}$. 

Next, we focus on the FM quantum critical behavior (QCB) in 
FeGa$_{3-y}$Ge$_{y}$. Although ferromagnetic or antiferromagnetic QCB 
has been observed in many $f$-electron systems,\cite{GR50} the FM QCB in 
$d$-transition metal systems has been identified on a much smaller number of 
compounds, such as ZrZn$_{2}$(Ref. 50) and 
Ni$_{x}$Pd$_{1-x}$.\cite{MN52} The FM QCB in these systems has been 
explained in terms of the SCR theory.\cite{TM47} For FeGa$_{3-y}$Ge$_{y}$, 
the experimental results of $C(T)$ and $M(T)/B$ near the critical concentration are 
consistent with the SCR theory of FM spin fluctuations, whereas the 
$T$-linear resistivity is at variance with the $T^{5/3}$ dependence predicted by 
this theory. Interestingly, Fe$_{0.7}$Co$_{0.3}$Si shows $T$-linear resistivity 
under the critical pressure of 7 GPa,\cite{YN11} whose origin of $\rho (T)$ is 
under debate. The resistivity is influenced by not only the spin 
fluctuations predicted by the SCR theory but also the band structure and 
disorder in the crystal. Therefore, an elaborate theory considering the 
actual band structure and the inherent effect of disorder is needed to 
explain the observed resistivity. Nevertheless, the electron correlation effect in FeGa$_{3}$ is 
not significant compared with FeSi,\cite{MA32} because of the absence of 
impurity induced density of states at the Fermi level indicated by the 
extremely small $\gamma $ value of 0.03 mJ/K$^{2}$mol.\cite{YH28} It is 
noteworthy that FeGa$_{3}$ with such a weak correlation effect exhibits the 
QCB near the critical point from the nonmagnetic state to the FM ground 
state. The QCB may be induced by strong spin fluctuations due to the 
disturbance in the Fe $3d$- Ga $4p$ hybridization. In order to clarify this point, 
neutron scattering studies on FeGa$_{3-y}$Ge$_{y}$ single crystals are 
highly desirable. 

\section{Conclusion}
The effect of electron doping on the electronic and magnetic states of a 
diamagnetic semiconductor FeGa$_{3}$ with a rather large band gap of 0.5 eV 
has been studied using single crystalline samples 
Fe$_{1-x}$Co$_{x}$Ga$_{3}$ and FeGa$_{3-y}$Ge$_{y}$. A 
semiconductor-metal transition in Fe$_{1-x}$Co$_{x}$Ga$_{3}$ occurs at 
$x = 0.23$, whereas no magnetic order is induced in the whole range $0 \le $ 
$x \le  1$. These observations can be explained by the gradual change of the 
band structure in the rigid-band frame. On the other hand, substitution of 
Ge for Ga in FeGa$_{3-y}$Ge$_{y}$ at a small value $y = 0.006$ yields 
metallic conduction, and further doping at $y = 0.13$ induces weak 
ferromagnetism. The $\gamma$ value as a function of $y$ exhibits a large peak 
of 70 mJ/K$^{2}$molFe at $y = 0.09$. The critical concentration $y_{c}= 
0.13$ for the ferromagnetism is rather small, in spite of the fact that the 
band gap of 0.5 eV is one order of magnitude larger than the gap sizes in 
FeSi and FeSb$_{2}$. The FM quantum critical behaviors are manifested as 
$C/T \propto  -$ln$T$ and $M/B \propto T^{-4/3}$ near the critical concentration of 
$y_{c} = 0.13$ in FeGa$_{3-y}$Ge$_{y}$. This FM instability is 
attributed to strong electron correlations, which are induced by the 
disturbance in the Fe $3d$ - Ga $4p$ hybridization by substituting Ge for Ga. 
Finally, we note that this system serves as a model system to investigate 
the FM instability in the simultaneous presence of disorder and electronic 
interaction, a problem that has been theoretically investigated.\cite{PB53}

\begin{acknowledgments}

We thank F. Iga and for fruitful discussions. The magnetization and specific 
heat measurements were performed at N-BARD, Hiroshima University. This work 
was supported by the Scientific Research (A) (18204032) from MEXT, Japan.  

\end{acknowledgments}


\newpage

\begin{figure}
 \includegraphics[width=12cm]{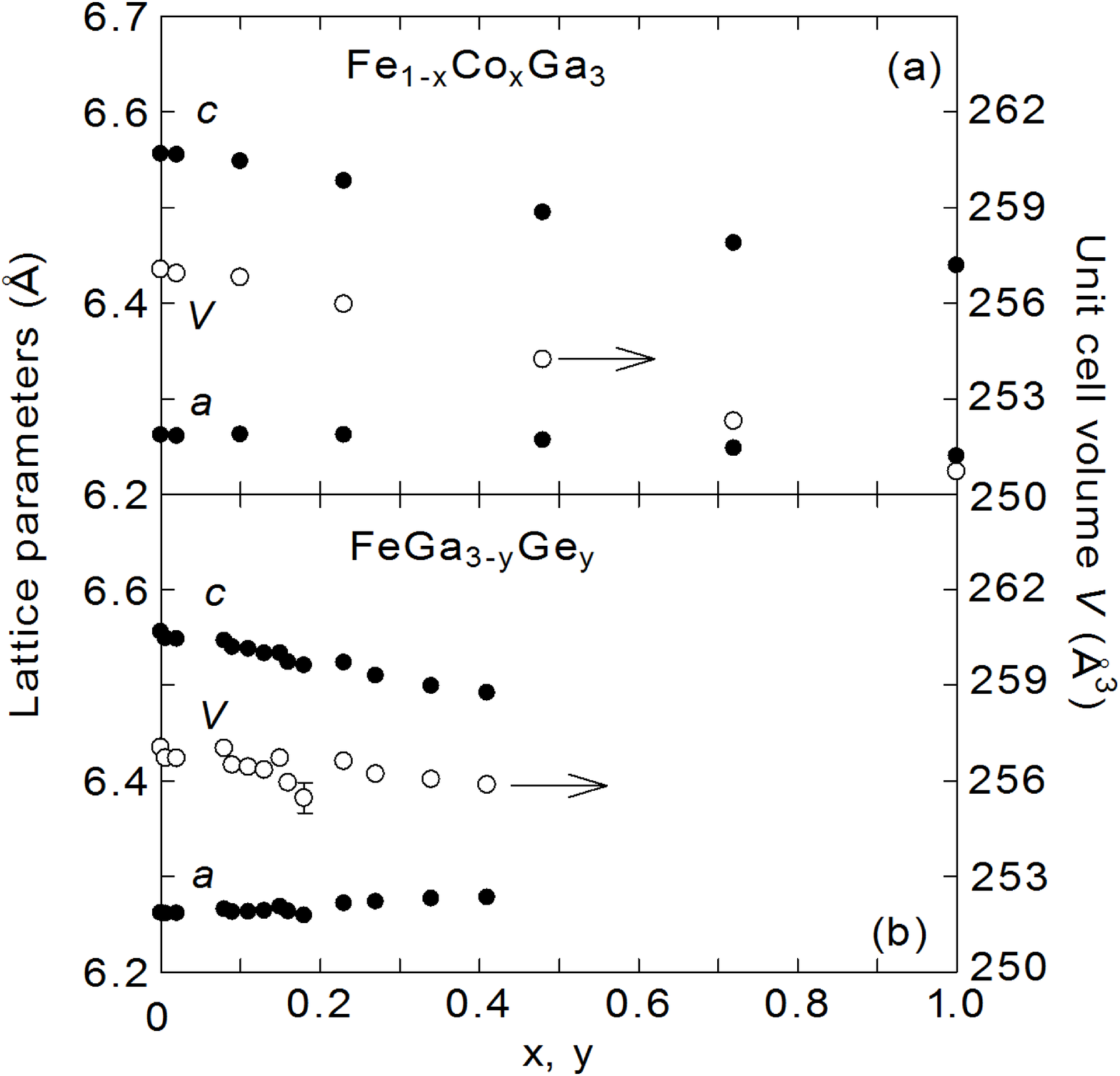}%
 \caption{\label{fig:epsart}Lattice parameters and unit cell volume of 
Fe$_{1-x}$Co$_{x}$Ga$_{3}$ (a) and FeGa$_{3-y}$Ge$_{y}$ (b) as a function of concentrations $x$ and $y$.
}
\end{figure}

\begin{figure}
 \includegraphics[width=10cm]{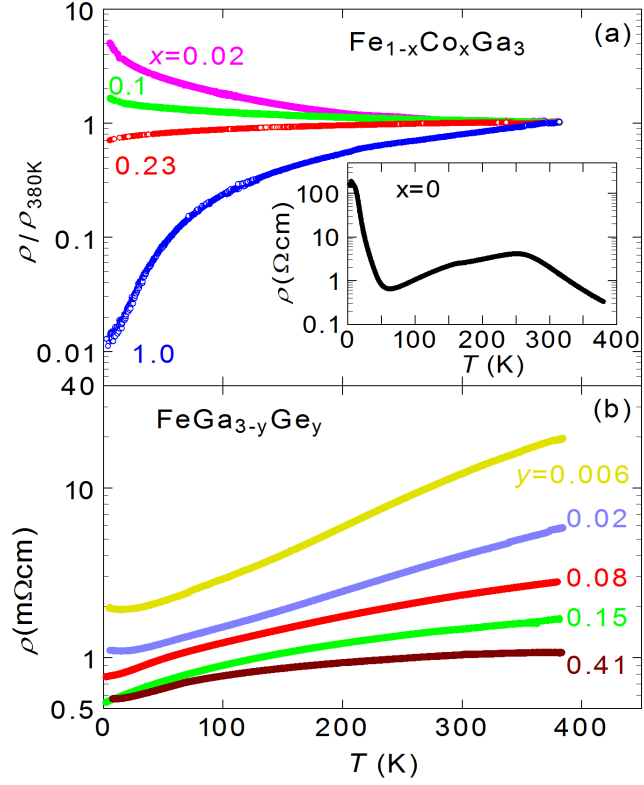}%
 \caption{\label{fig:epsart}Temperature dependence of electrical resistivity $\rho$ for 
Fe$_{1-x}$Co$_{x}$Ga$_{3}$ (a) and FeGa$_{3-y}$Ge$_{y}$ (b). The 
resistivity of Fe$_{1-x}$Co$_{x}$Ga$_{3}$ is normalized by the value at 
380 K. The inset shows the resistivity for FeGa$_{3}$ ($x = 0$).\cite{YH28}
}
\end{figure}

\begin{figure}
 \includegraphics[width=10cm]{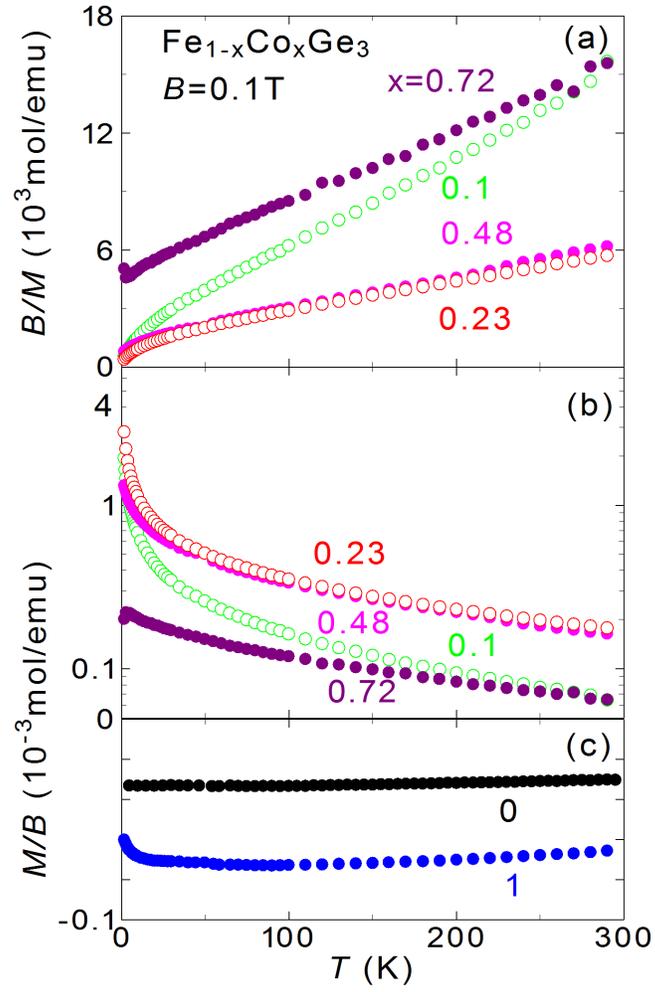}%
 \caption{\label{fig:epsart}Temperature dependence of magnetic susceptibility $M/B$ (b) and inverse 
magnetic susceptibility $B/M$ (a) of Fe$_{1-x}$Co$_{x}$Ga$_{3}$ 
}
\end{figure}

\begin{figure}
 \includegraphics[width=10cm]{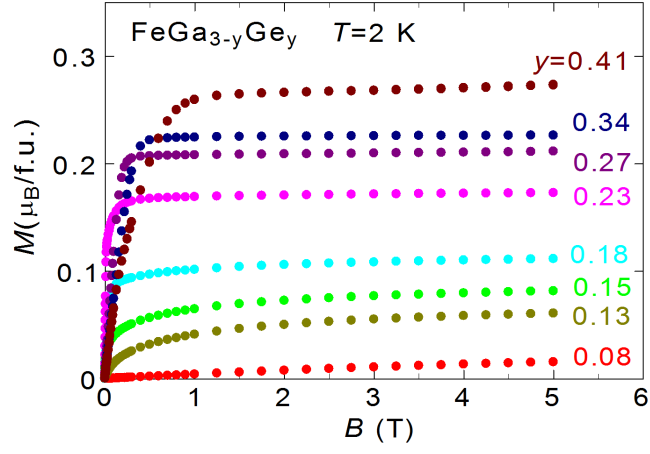}%
 \caption{\label{fig:epsart}Isothermal magnetization curves of FeGa$_{3-y}$Ge$_{y}$ at 2 K.
}
\end{figure}

\begin{figure}
 \includegraphics[width=10cm]{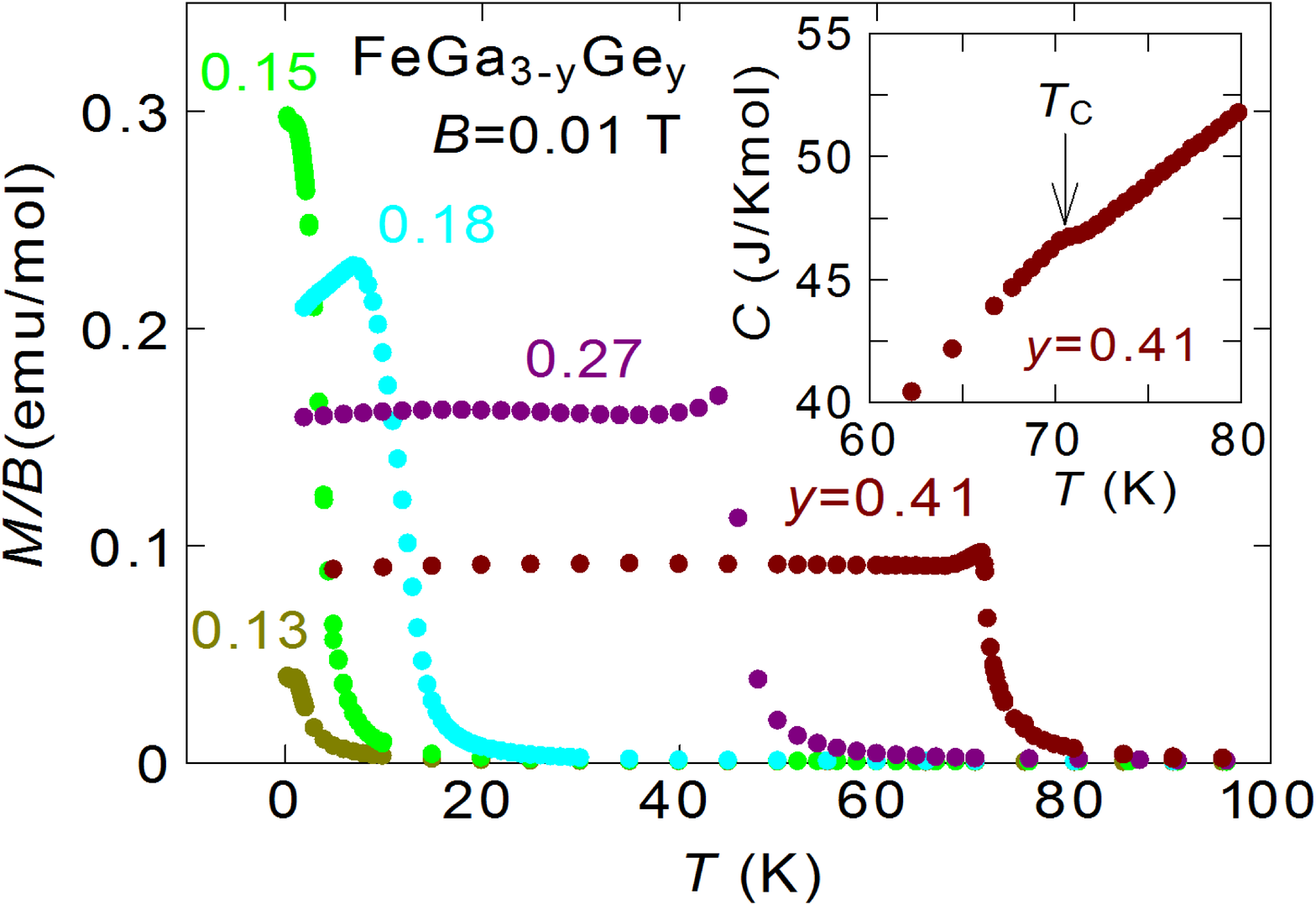}%
 \caption{\label{fig:epsart}Temperature dependence of magnetic susceptibility $M/B$ of 
FeGa$_{3-y}$Ge$_{y}$ for $y \ge  0.13$ where ferromagnetic transitions 
are observed. The inset shows the specific heat of FeGa$_{3-y}$Ge$_{y}$ for $y = 0.41$ near $T_{C}$.
}
\end{figure}

\begin{figure}
 \includegraphics[width=10cm]{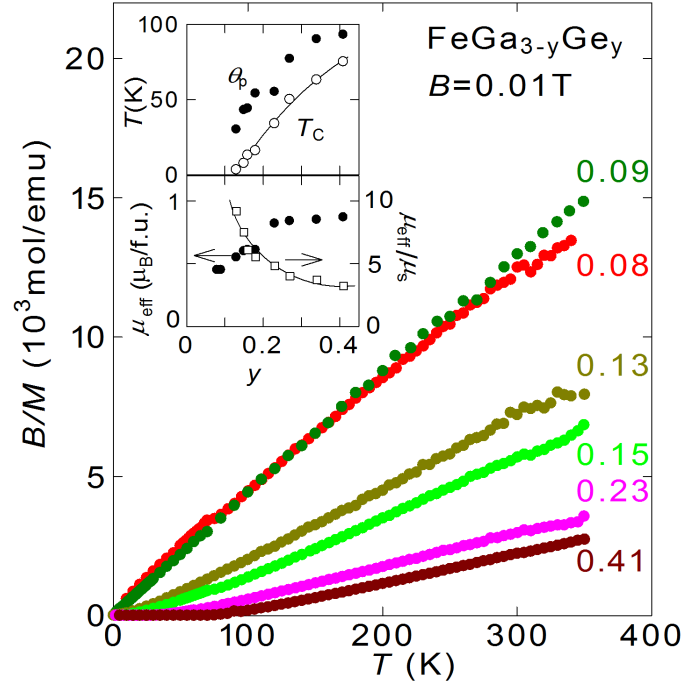}%
 \caption{\label{fig:epsart}Temperature dependence of the inverse magnetic susceptibility $B/M$ of 
FeGa$_{3-y}$Ge$_{y}$. The upper and lower panels of the inset show the 
paramagnetic Curie temperature $\theta_{P}$ and ferromagnetic transition 
temperature $T_{C}$, and effective magnetic moments $\mu_{eff}$ and the 
Rhodes-Wohlfarth value $\mu_{eff}$/$\mu_{s}$, respectively, as a 
function of $y$.
}
\end{figure}

\begin{figure}
 \includegraphics[width=10cm]{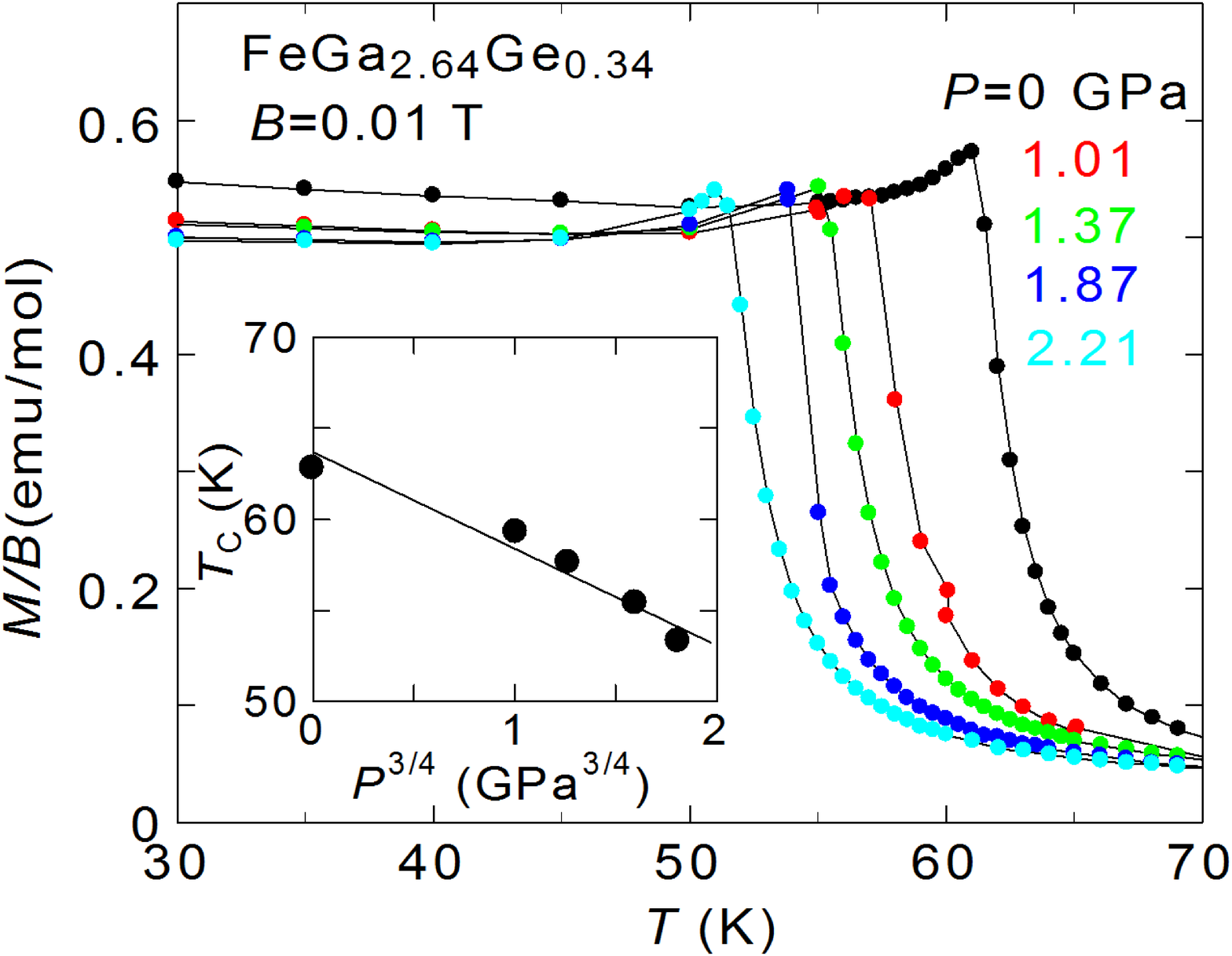}%
 \caption{\label{fig:epsart}Temperature dependence of magnetic susceptibility $M/B$ of 
FeGa$_{3-y}$Ge$_{y}$ for $y = 0.34$ under various pressures $P$. The inset 
shows $T_{C}$ as a function of $P^{3/4}$.
}
\end{figure}

\begin{figure}
 \includegraphics[width=10cm]{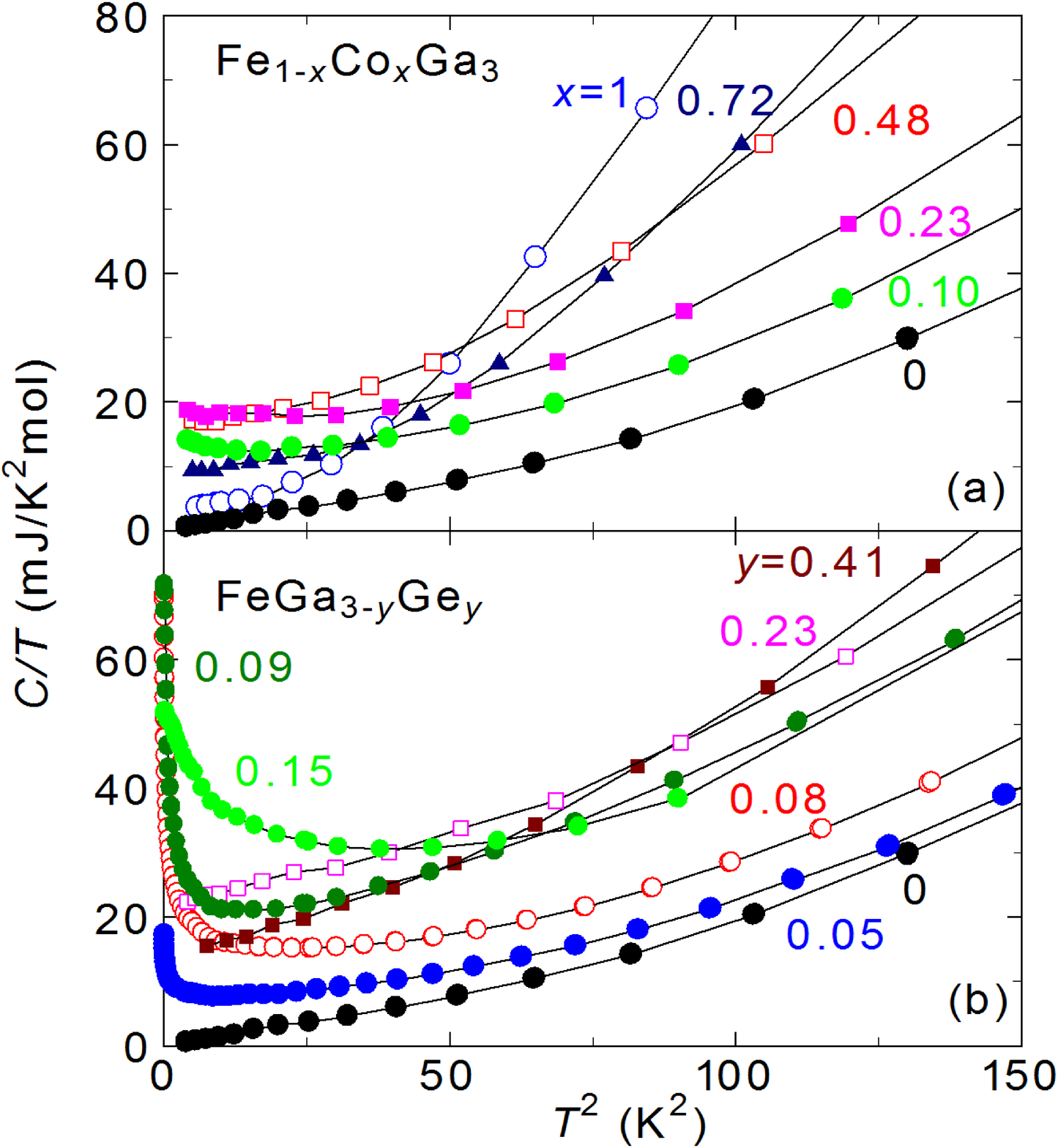}%
 \caption{\label{fig:epsart}The specific heat divided by temperature $C/T$ for 
Fe$_{1-x}$Co$_{x}$Ga$_{3}$ (a) and FeGa$_{3-y}$Ge$_{y}$(b) as 
a function of $T^{2}$. 
}
\end{figure}

\begin{figure}
 \includegraphics[width=12cm]{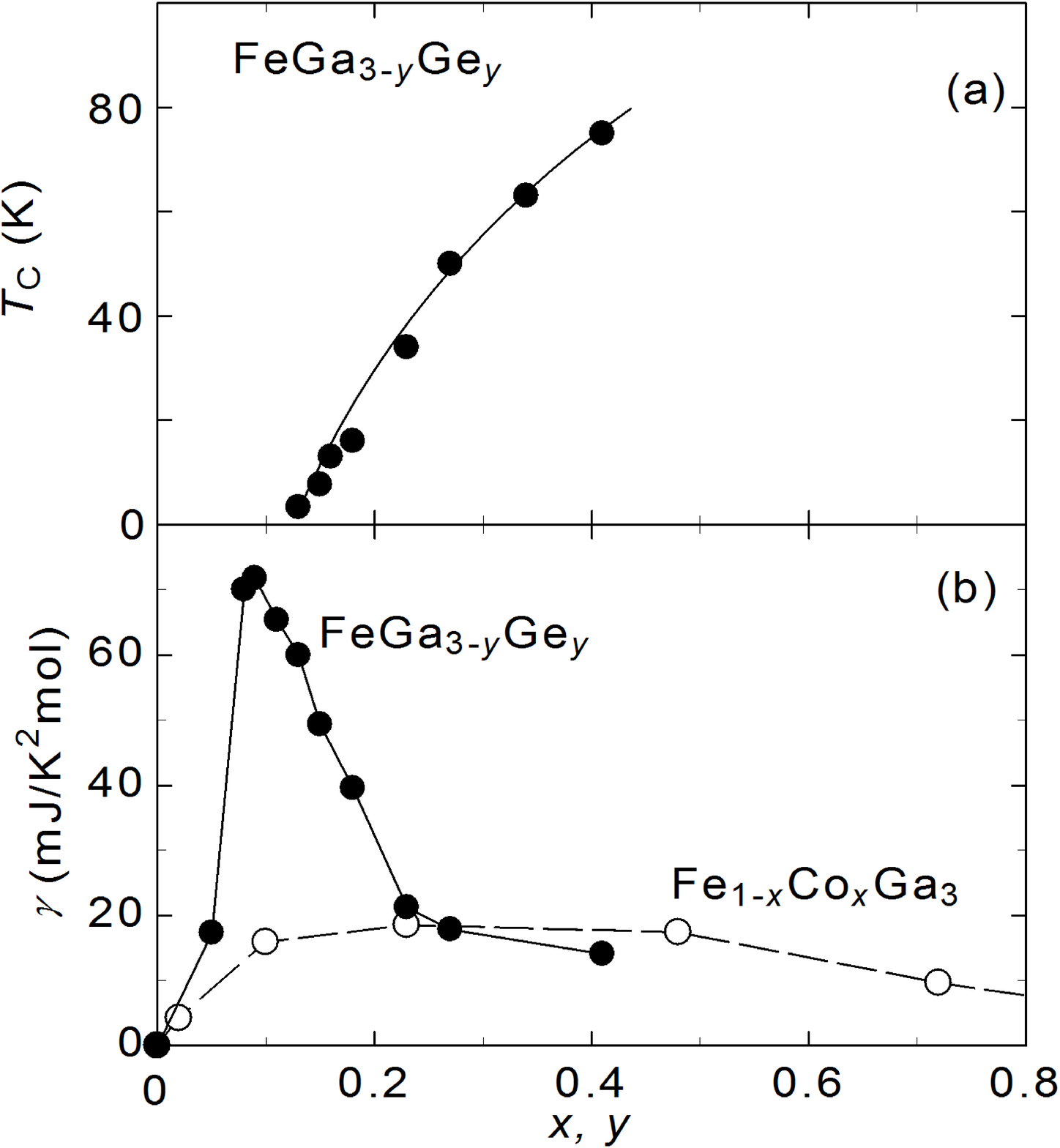}%
 \caption{\label{fig:epsart}Ferromagnetic transition temperature $T_{C}$ (a) and electronic 
specific heat coefficient $\gamma $ (b) for Fe$_{1-x}$Co$_{x}$Ga$_{3}$ and FeGa$_{3-y}$ Ge$_{y}$ as a function of $x$ and $y$.
}
\end{figure}

\begin{figure}
 \includegraphics[width=12cm]{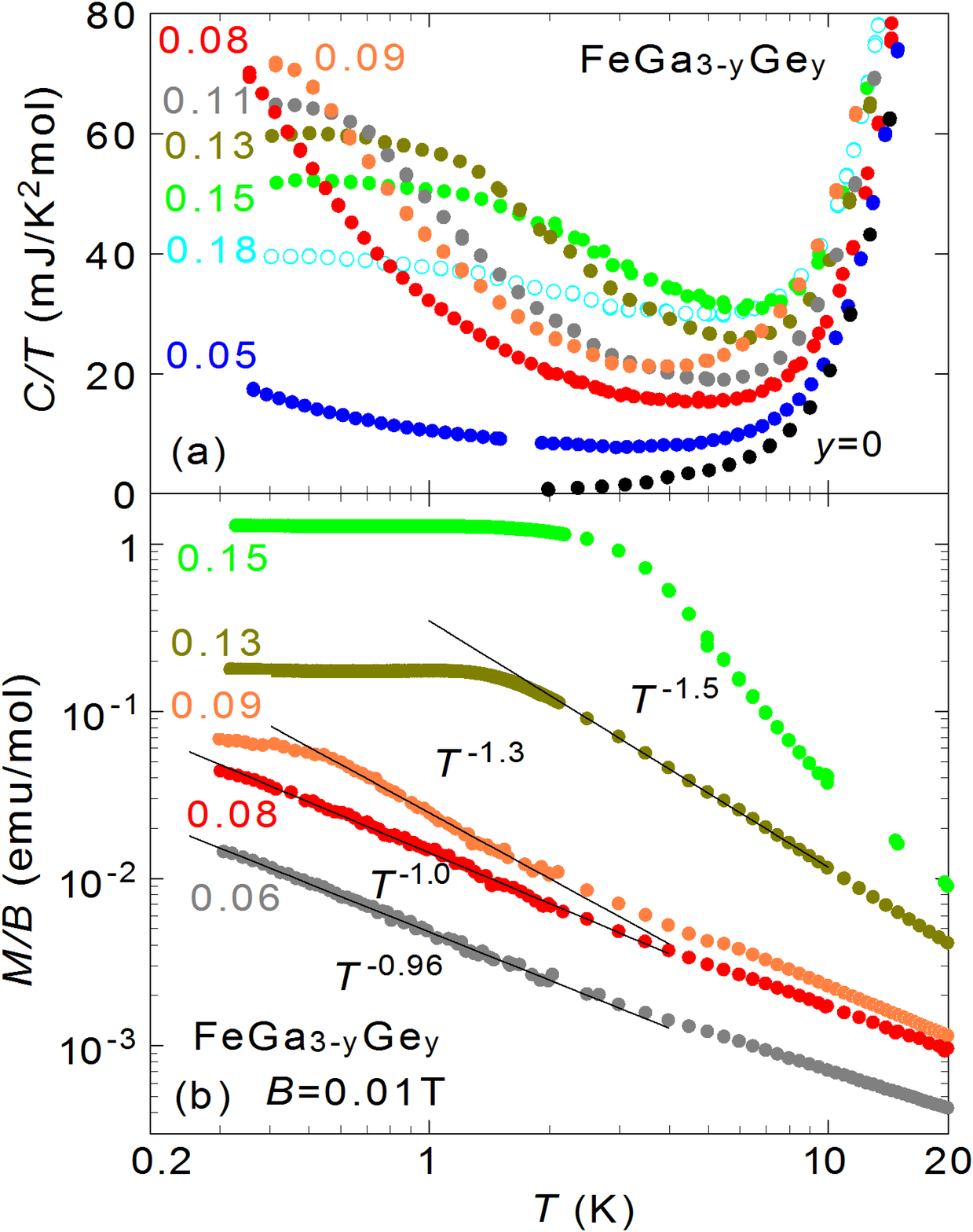}%
 \caption{\label{fig:epsart}Logarithmic temperature dependence of $C/T$ (a) and $M/B$ (b) for 
FeGa$_{3-y}$Ge$_{y}$ near the FM instability.
}
\end{figure}

\begin{figure}
 \includegraphics[width=10cm]{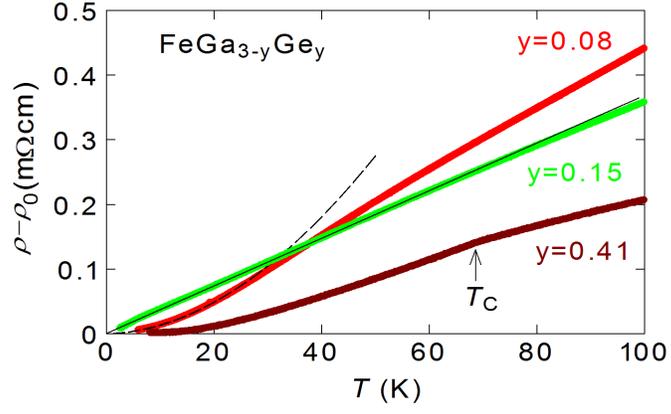}%
 \caption{\label{fig:epsart}Temperature dependence of electrical resistivity $\rho$ of $y = 
0.08$, 0.15, and 0.41 for FeGa$_{3-y}$Ge$_{y}$. The $\rho (T)$ data for 
$y = 0.08$, and 0.15 were fitted by $\rho -\rho_{0} \propto T^{1.9}$ 
(broken line) and $\rho -\rho_{0} \propto T$ (solid line), 
respectively. 
}
\end{figure}

\end{document}